# Decision-making processes underlying pedestrian behaviours at signalised crossing: Part 1. The first to step off the kerb.

Marie Pelé[1], Jean-Louis Deneubourg[2], Cédric Sueur[3]

[1] Ethobiosciences, Research and Consultancy Agency in Animal Well-Being and Behaviour, Strasbourg, France

[2] Unit of Social Ecology, Université libre de Bruxelles, Brussels, Belgium

[3] Université de Strasbourg, CNRS, IPHC UMR 7178, F-67000 Strasbourg, France

Corresponding author: marie.pele@iphc.cnrs.fr, 0033(0)88107453, IPHC UMR 7178, 23 rue Becquerel F-67000 Strasbourg, France

**Abstract:** Human beings have to make numerous decisions every day, and these decisions might be biased and influenced by different personal, social and/or environmental variables. Pedestrians are ideal subjects for the study of decision-making, due to the inter-individual variation in risk taking. Many studies have attempted to understand which environmental factors (light colour, waiting times, etc.) influence the number of times pedestrians broke the rules at road-crossings, very few focused on the decision-making process of pedestrians according to the different conditions of these variables, that is to say their perception and interpretation of the information they receive. This study used survival analyses to highlight the decision-making process of pedestrians crossing the road at signalized crossings in France and in Japan. For both light colours, we decided to carry out separate analyses for the first pedestrian to step off the kerb and other individuals following him/her, as the decisions underlying the departure of a first individual and those of the followers are different, and the departure of the first pedestrians strongly influences the decisions of other individuals. We showed that the probability to cross the road follows three different processes: one at the red light, one just before the pedestrian light turns green, and one after the light has turned green. Globally, the decision of the first pedestrian to cross, whether he or she does so at the green or at the red light, is influenced by their country of residence. We observed a lower threshold for Japanese pedestrians because they tend to follow their private or personal information. We identify the use of cognitive processes such as risk sensitivity and temporal discounting, and propose new concepts based on the results of this study to decrease the incidence of rule-breaking by pedestrians.

**Introduction**

Human beings have to make numerous decisions every day throughout their lifetime. Most of these decisions are relatively easy: what to eat for breakfast, what to wear, what itinerary they will use to go to school or to work. Whether or not they should marry or have a child are bigger choices to make. Finally, some individuals can change the lives of millions of people with their decisions to vote laws or go to war. Many studies have attempted to understand how decisions are taken and if they are optimal from an evolutionary perspective [1,2]. Decisions usually follow something comparable to an optimal test called SPRT (sequential probability ratio test) and require a sufficient difference of evidence or information between two alternatives in order to choose the most profitable or the less risky of the two options [3,4]. However, without going so far as to claim that these everyday life decisions are suboptimal [5,6] or irrational [7–9], human decisions might be biased and influenced by personal, social and/or environmental variables.

Pedestrians are ideal subjects when studying decision making. Pedestrians need to perceive and integrate a great deal of information compared to other situations they encounter. They have to identify spatial cues about where to go, avoid other pedestrians as they walk in the street [10,11] and cross the roads, which can sometimes be a high-risk behaviour [12,13]. Like in other behaviours, pedestrian behaviour -and particularly road crossing - displays great variance which is dependent on many factors [12,14–18]. Some pedestrians will take more risks than others when crossing the road, either by crossing at the red light, or by decreasing the gap acceptance with a car [19,20]. Indeed, a higher rate of risky behaviours has been observed in males or young individuals, particularly adolescents [18,21,22]. Older persons have sometimes been reported to take more risks than others, but this is due to a loss of perceptive and cognitive abilities rather than intentional risk taking [13]. Some people also prefer to base their decision on their personal information instead of trusting the social information [23,24]. This might be particularly advantageous in the case of road crossing, i.e. choosing to not follow pedestrians crossing at the red light without checking how far away the next car is [25,26]. The use of social information and the probability of rule breaking are strongly correlated with the culture and the country of pedestrians, with each country having its own principles of conformism and social norms [27,28]. Previous studies have shown that the number of illegal crossings is largely dependent on the country where pedestrians live, and its culture [26,29,30]. However, the majority of these studies made correlational analyses but did not explore the possible presence of cognitive mechanisms underlying the decision-making processes [3,4].

Indeed, whilst many studies tried to understand which factors influence the incidence of rule breaking at road crossings, very few focused on the decision-making process of pedestrians facing the different conditions of these variables, that is to say how their perception and interpretation of the information they receive [31]. This study aimed to highlight the decision-making process of pedestrians crossing the road at a signalized crossing. For both light colours, we decided to separate analyses between the first pedestrian to go and other individuals following him/her, and focused the first part of our study on the first pedestrian to step off the kerb. Indeed, the decisions underlying the departure of a first individual those of the followers are different, and the departure of the first pedestrians strongly influences the decisions of other individuals [32,33]. The first pedestrian to step off the kerb is also the only one to exclusively follow his/her own personal information, without being influenced by other pedestrians, which could be considered a high-risk behaviour.

This study used survival analysis to understand decision-making processes. Survival analysis is a statistical tool used to predict when one or more events will occur, such as death in biological organisms and failure in mechanical systems [34–36]. In the case of pedestrians crossing the road, survival analysis allows us to understand how the probability that pedestrians will cross is influenced by time according to the imminence of the pedestrian light changing, whether from red to green or *vice versa*. The decrease of the curve provides information about the process underlying the decision, with a sigmoid indicating a threshold similar to that of a diffusion model, whilst an exponential decrease shows that the probability is constant per time unit [2,26,37]. Survival analysis can also be used to test other factors such as the gender or culture of pedestrians, the number of traffic lanes or the number of waiting pedestrians. In this study, these factors are analysed in order to identify time thresholds to cross the road and pinpoint optimal decision processes involved in pedestrian road-crossing. However, thresholds should differ according to the perceived risk [38]: for instance, it should be lower in men than in women. Similarly, we expected thresholds to increase with the number of lanes, as the perceived risk is higher in this situation. The effect of culture on pedestrian road-crossing behaviour was studied in sites in France (Strasbourg) and in Japan (Nagoya). We expected to see a difference between the two countries, with the threshold for Japanese pedestrians occurring closer to the time the light turned green. This is not due to higher perceived risk, but is rather explained by their known conformism to rules, contrary to pedestrians in Western countries [39].

**Material & Methods**

    a. Study sites

We observed pedestrian behaviours at three sites in Strasbourg, France and at four sites in Nagoya,

Japan. Details about each site are given in Table 1. Pictures of the different sites are available in [26]. These sites all permitted the observation of collective road crossings involving at least 10 pedestrians at a time. The speed of vehicles on each site was limited to 50kmh$^{-1}$. There was no difference in pedestrian crossing speed between the sites (permutation test for independent samples: maxT=2.22, p=0.168). At some sites, vehicles were allowed to turn left or right despite the green light for pedestrians, but the drivers were aware that crossing pedestrians had priority. Moreover, turning vehicles travel much slower than vehicles that are driving straight ahead. However, the driver of an approaching vehicle may be less careful if pedestrians cross at the red light, as the driver has the right to pass. The risk to pedestrians is therefore much higher when crossing at the red light. There was no button for pedestrians to trigger the green pedestrian light at any of the sites studied.

b. Data scoring

Data were scored over a 6-day period for each site, for 1 h per day during working days, hours and weeks to ensure that data excluded movements generated by tourism, festivals, etc. This scoring duration is sufficient to provide a large dataset [2,20,25]. Video cameras were set up in order to score the light colour and were placed in locations ensuring the visibility of crossing pedestrians at all times. Behavioural sampling was used to score the crossing of pedestrians in one direction only, i.e. that recorded by the camera. Pedestrians were not informed about the purpose of the study. As both cities are touristic, pedestrians are accustomed to seeing tourists taking pictures or videos. We did not observe any difference in the way pedestrians behaved when they saw the camera. We purposely did not take any other equipment such as counters or pocket PCs in order to avoid influencing pedestrian behaviour. When the observation of road-crossing behaviour was hampered by a visual obstacle (i.e. a car or a truck in front of the video camera), this behaviour and the behaviours occurring immediately before and after it were removed from the data set, as were any data recorded when cyclists or tourists were among the pedestrians. Tourists were easily differentiated from local citizens, as they were dressed differently from citizens, often carried specific equipment (guidebook, map, camera, etc.), and/or were in large groups accompanied by a guide.

a. Research ethics

Our methodological approach solely involved anonymous observations and anonymous data scores. Study protocol followed the ethical guidelines of our research institutions (IPHC, Strasbourg, France and PRI, Kyoto University, Japan) and ethical approval was obtained from these institutions to carry out the study. All data were anonymous, and individuals were given sequential numerical identities according to the time of the road crossing and the arrival/departure order of crossing. Pedestrians had the possibility to obtain information about the study via a fact sheet in their language (Japanese or French). They were also provided with an email address and phone number to contact our institution at a later date if desired. Persons who refused to participate in the study were removed from the data (i.e. we deleted the crossing concerned).

b. Data analysis

This part one of the study focused solely on the first pedestrian to step off the kerb (at the red or green light) and not on following pedestrians (see part 2. Pelé et al. submitted), because the processes underlying the two decisions (departing first and following) are quite different [40–42]. However, we also selected cases where the first pedestrian to step off the kerb is in the presence of other pedestrians, in order to understand the impact of this variable on decision-making processes. All 6 h of data were analysed for each site. We scored the behaviours of the first pedestrians to step off the kerb when at least two pedestrians crossed the road at the same time (i.e. when the time difference between the departures of the two pedestrians was lower than the mean road-crossing time indicated in Table 1 for each site).
We scored road crossings for 429 first pedestrians, 244 of whom crossed at the green light and 185 of whom crossed at the red light.

For each first pedestrian to step off the kerb, we scored the following variables (see [26] for a visual explanation of the different scored variables):
— The light colour when crossing (red or green).
— The departure period, i.e. the period between the previous light colour change and the moment the pedestrian starts crossing the road. This variable is positive for pedestrians crossing at the green light (after the colour change) but negative for pedestrians crossing at the red light (before the colour change).
— The gender of pedestrians (male or female).
— The age of individuals, estimated at 10-year intervals from 0–9, 10–19 [. . .] to 70–89. However, the number of data and the analyses we carried out did not permit the analysis of age effect (per interval) on the decision-making processes.
— The country (France or Japan).
— The number of road lanes.
— The total number of waiting pedestrians.

— The waiting time, i.e. the time between the moment a pedestrian stops at the light and the moment he/she starts crossing the road.

c. Statistical analyses

Survival analysis [35,36] was used to study the distributions of departure periods for the first pedestrian to step off the kerb. Survival analysis is used to understand how the ratio of observations decreases from 1 (all observations/data) to 0 (none) according to a response variable. First, curve estimation tests were carried out to analyse which type of function these distributions followed, namely linear (meaning that the probability of crossing depends directly on time), exponential (the probability of crossing is time constant) or sigmoid (the probability of crossing depends on a time threshold that is directly correlated to the response variable, see [13,37,43]).

Sigmoid curves (Equation 1 and Equation 2) are generally used to understand decision-making processes [37]. Other curves used for these studies are the linear curve (Equation 3, meaning that probability of crossing depends directly on time, regardless of factors such as the distance to the next car or the number of waiting pedestrians), and the exponential curve (Equation 4, where the probability of crossing is time-constant; see [13]. Sigmoid curves are indicated with two parameters, S and q. q is a sensitivity coefficient. In essence, a higher q value results in a faster transition between resting and departing and therefore also results in higher discrimination [37]. S is a threshold. The higher it is, the longer it will take to reach a decision. In the case of road crossing, the threshold S would be more representative of the risk taken by individuals, whilst q would be more dependent on cognitive aspects (namely perception and interpretation of information) and individual traits [13]. If the distribution of the departure times corresponds to an exponential distribution, the departure probability of the first pedestrian to step off the kerb is the log gradient $b$ of the corresponding exponential distribution, i.e, the inverse of the mean departure time $\left(\frac{1}{\Delta t}\right)$ if Δt is the mean departure time.

Equation 1: $y = \frac{1}{1+e^{q\frac{x}{S}}}$

or Equation 2: $y = \frac{1}{1+e^{\frac{1}{q}*\frac{x}{S}}}$

Equation 3: $y = ax + b$

Equation 4: $y = a * e^{-bx}$

Where y is survival and x is the departure time or the studied variable.

We also simulated the crossings of pedestrians at the red light under the hypotheses that the probability to arrive at the kerb and the probability to cross at the red light is constant. These simulations allowed us to compare our observed data to these theoretical data. The survival curve NC(T) following these simulations is:

$$NC(T) = e^{-kT}(1 - \frac{T}{F})  \qquad \text{equation 5}$$

With F is the time of the red light; k is the probability to cross at the red light; T is the waiting time.

We tested two k values (k=0.0005, k=0.05). Results of these simulations are shown in fig.2.

Linear regression was used to analyse the distribution of observed data by comparing it to the distribution of theoretical data with adjusted $R^2$. Fitting distributions (linear, sigmoid and exponential) were chosen according to F-statistics. Differences found in the equational parameters between the countries and between the genders were tested using a Wilcoxon sign rank test. Levene tests were carried out on transformed data (log(Survival/b) for exponential and Ln((1/Survival)-1) for sigmoid curves) to compare departure times according to the number of lanes and the country. The same approach was applied to compare the waiting times according to country and gender. Analyses were performed in R 3.3.2, with α set at 0.05. Sequential Bonferroni correction was used [44,45] for multiple variables analyses. However, given the p-value of our tests, this did not change their significance.

**Results**

When all data has been analysed (red and green lights), the distribution of the departure time on Figure 1 is far from perfect as a sigmoid curve, despite being significant ($R^2$=0.63, df=427, p<0.0001, F=3069, q=0.3 and S=73). This reveals that at least two different rules underlie a decision to step off the kerb for the first pedestrians. We then decided to carry out separate analyses of instances where pedestrians departed first at the red light, and those where pedestrians departed first at the green light.

    a. Analyses of departure times at the red light

Analyses of departure times at the red light showed a reliable estimation of observed data using a sigmoid law (Equation 2, R²=0.87, df=173, p<0.0001, F=1146, q=10, S=-100). Times of departure follow an asymptote, with a plateau decreasing faster and faster as time approaches 0 (i.e. when the light turns green). If the probability of departures would be constant per time unit, the survival would be either linear (orange points, k=0.0005, small probability to cross at the red light) or exponential-like (blue points, k=0.05, high probability to cross at the red light). So our results show that the probability to cross at the red light is not constant but is almost null when the pedestrian light goes red and then increases when getting closer and closer to the green pedestrian light. Indeed, the plateau we observe from observed data seems to show a kind a refractory phase to depart after the pedestrian light goes red. However, Fig.2 shows different drops in the two asymptotic curves (grey and yellow), suggesting two different processes. The two different curves have a breaking point at -400 $sec^{-100}$; we consequently used this departure time as a marker to divide our analysis of road crossings at the red light.

Analyses of departure times at the red light before -400 $sec^{-100}$ seconds showed a reliable estimation of observed data using a sigmoid law (Equation 2, R²=0.98, df=88, p<0.0001, F=5208, q=3, S=-300, Fig.3a,b). Times of departure follow an asymptote, with a plateau that decreases faster and faster with the x axis as we approach the -400 $sec^{-100}$ marker. Analyses of departure times at the red light between -400 $sec^{-100}$ and 0.00 showed a reliable estimation of observed data using a sigmoid law (Equation 2, R²=0.99, df=93, p<0.0001, F=11040, q=5, S=-180, Fig.3c,d).

b. Analyses of departure times at the green light

The time between the light turning green and the departure of the first pedestrian (departure time) follows a sigmoid curve (R²=0.98, df=242, p<0.0001, F=15670, Equation 1, S=73, q=2.5, Fig.4). No further analysis is required at this stage to understand time before departure at the green light.

c. Effect of country and gender on the time of departure of the first pedestrian

Here, we used the same procedure as in the previous analyses to evaluate the best parameters of the sigmoid curve explaining the distribution (survival) of the departure time of pedestrians in four categories: man from France, woman from France, man from Japan, and woman from Japan (whatever their age).

The threshold and sensitivity coefficients for each of these four categories are indicated in Table 2. We could not determine the threshold and the sensitivity coefficient for departure times at the red light (<-400 $_{sec^{-100}}$) in male and female Japanese pedestrians, as only two values were obtained per category, representing 1.47% of total crossings for Japanese men and 0.04% for Japanese women. This is already a result in itself, as the percentage of crossings at the red light (<-400 $_{sec^{-100}}$) for French men and women are 43.01% and 54.94%, respectively. Japanese pedestrians, whatever the sex, have a threshold closer to 0, i.e. the time at which the light turns green (Sign rank test, v=21, p=0.03) - a stark difference with French pedestrians. This is not the case with the sensitivity coefficient (sign rank test, v=4, p=0.85). There is no difference between men and women, whatever the country, for the threshold (Sign rank test, v=5, p=0.422) or the sensitivity (sign rank test, v=1, p=0.197).

d. Effect of the number of lanes

We then tried to understand the effect of the number of lanes on the probability that a pedestrian would cross. No rule breaking was observed at the site with six lanes in Japan. This is already a result, showing that the number of lanes impacts the probability of crossing at the red light. Survival curves for crossing at the green light are sigmoid (2 lanes: $R^2$ = 0.98, df=2133, p<0.0001, F=18614; 4 lanes: $R^2$ = 0.99, df=106, p<0.0001, F=10010), as found in the first part of results with the same sensitivity (q = 3.5). However, the threshold is higher (S = 90 $_{sec^{-100}}$) for crossings on roads with four lanes than those with two lanes (S = 67 $_{sec^{-100}}$). Survival curves are exponential for crossings at the red light (2 lanes: $R^2$ = 0.98, df=114, p<0.0001, F=5765; 4 lanes: $R^2$ = 0.98, df=65, p<0.0001, F=3468), indicating that the probability of crossing is constant per time unit. The exponents of these curves make it possible to calculate the average of the departure times at which individuals crossed the road (see Material & Methods), namely -1666 $_{sec^{-100}}$ (1/Δt = 0.0006) for 2 lanes and -285 $_{sec^{-100}}$ (1/Δt = 0.0035) for 4 lanes. When we consider crossings at the red light, the number of lanes and the country both affect the probability that a pedestrian will cross (Table 3). Japanese pedestrians cross closer to the light change compared to their French counterparts (Levene test, df=1, 176, F=15.291, p=0.0001), and the presence of 4 lanes on roads decreases the probability to cross at any time (Levene test, df=1, 176, F=4.59, p=0.033).

e. Effect of the number of waiting pedestrians

When attempting to understand how pedestrians decide to cross according to the number of waiting pedestrians, we found that survival curves at both the red and the green light follow sigmoid curves (Green light: $R^2$ = 0.99, df=242, p<0.0001, F=32490, S = 11, q = 5; Red light: $R^2$ = 0.99, df=183,

p<0.0001, F=30070, S = 11, q = 3.5; Fig 5a). However, with the exception of a sensitivity coefficient which seems to be lower at the red light, the two curves are quite similar (Levene test: df=1, 425, F=0.0034, p=0.953). This might be explained by the possibility of similar processes underlying the decision to step off the kerb. These may be linked to the number of waiting pedestrians, or to the number of observations decreasing according to the number of pedestrians waiting for the green and the red light. The latter situation would result in a lower probability of numerous waiting pedestrians, which is quite understandable. To check this hypothesis, another analysis was carried out to measure the ratio $\frac{\text{number of observations at the green light} - \text{number of observations at the red light}}{\text{total number of observations}}$ for each number of waiting pedestrians during road crossings at the red light and at the green light (Fig.5b). The best curve explaining the distribution of data is a cubic curve ($R^2 = 0.22$), meaning that we observed a higher rate when there were fewer people waiting. However, the regression analysis is not significant. Only the first part of the graph (red square going from 1 to 11 waiting pedestrians) shows a good fit between observed data and the theoretical curve ($R^2 = 0.97$), whilst the remaining numbers of waiting pedestrians display huge variations in the ratio of observations.

f. Effect of waiting time

Waiting time, meaning the time between the arrival of a pedestrian at the kerb and his or her departure, followed an exponential curve (Green light: $R^2 = 0.94$, df=222, p<0.0001, F=3670; Red light: $R^2 = 0.99$, df=179, p<0.0001, F=15710). This time is not significantly different according to whether the pedestrian starts at the green or at the red light (Levene test: df=1, 403, F=0.0021, p=0.942). When considering crossings at the red light alone (at the green light, the time is influenced by the light change), the survival curve also follows an exponential law regardless of country and gender (Table 4). Levene tests did not reveal any significant difference between genders (Levene test: df=2, 175, F=0.047, p=0.953) and countries (Levene test: df=1, 176, F=0.293, p=0.589).

**Discussion**

In this study, we tried to understand the decision-making processes underlying road crossing behaviours at a signalised crossing. Survival analyses not only show whether variables (light colour, gender, country, number of waiting pedestrians and number of lanes) impact the way pedestrians cross, but also how pedestrians integrate this information in their decision to cross.

The results show that the probability to cross the road follows three different processes: one at the red light up to 400 sec$^{-100}$ (or four seconds) before the light turns green, one between 400 and 0

sec$^{-100}$, before the light turns green, and one after the light has turned green. These three processes are easy to explain. The first process corresponds to pedestrians who do not pay attention to the light colour and have little fear of risk-taking (or at least are more inclined to this attitude than pedestrians following the two other processes). The last process (crossing at the green light) corresponds to people who pay attention to the light colour and do not take risks. The intermediary process we found might be explained by pedestrians who cross the road just before the light turns green. The curve analysis of the first process (before -400 sec$^{-100}$) simply shows an incidence of rule-breaking that increases as the green light change approaches, or inversely, an incidence of rule-breaking that becomes rarer and rarer as the green light change approaches.

There are two possible explanations for the first process (crossing before -400 sec$^{-100}$). The first reason is that early departure times (-70 or -60 seconds before the green light) are rare because the flow of cars at this time is dense due to the green light for cars. As the time to the next pedestrian green light decreases, the flow of cars diminishes but the number of pedestrians waiting at the kerb increases. The second possible explanation of this process is the probability of seeing a pedestrian departing as a consequence of the increasing number of pedestrians waiting to cross. The increasing number of pedestrians increases the probability of seeing one pedestrian crossing. However, the waiting time of pedestrians also increases their probability to cross. Studies have shown that the longer people wait, the higher the probability is that they will cross at the red light [46,47]. We observed an effect of country on this rule breaking, with very few pedestrians in Japan crossing at the red light when it is not close to changing, which is reminiscent of previous studies on the effect of culture on decision making [13,26]. The current results confirm this study with no difference in the distribution of waiting times at the red and the green light, meaning that pedestrians do not seem to plan crossing at the red light: they arrive and wait, but will cross illegally if the waiting time is too long.

Our study identifies an intermediary process, which is quite different to the processes of crossing at the red light and crossing at the green light. Indeed, pedestrians crossing just before the light turns green checked if any car was arriving, or crossed because the light for cars had turned red a few seconds before the pedestrian light turned green. This step-by-step light change is present in most countries, if not all, in order to decrease the risks of accidents between cars and pedestrians, mainly because pedestrians need time to cross the road, especially in the case of old or disabled persons [48]. The survival curve of this process is closer to a linear law than to an asymptote. Usually, linearity in survival analysis indicates that the probability we measured is time dependent [13,49]. In our case, this might be due to pedestrians seeing the risk decreasing with time (Step 1: amber light for cars, step 2: red light for cars, step 3: green light for pedestrians) and the decision changes over time. This process is called temporal discounting and has already been described in humans beings in different situations [2,50].

The last process, i.e. crossing at the green light, perfectly follows a sigmoid curve. Finding a sigmoid curve here and not an exponential law means that some type of cognitive processes, presumably for decision-making, underlie the choice of departure time [2,3]. In this kind of process and according to the diffusion model (Bogacz, 2007), individuals need to obtain enough information to take an optimal decision. This shows a speed-accuracy trade-off, and involves a threshold for which an alternative (in this case, crossing or continuing to wait) is chosen. The threshold, set here at 73 $sec^{-100}$, shows the necessary time to obtain sufficient information between the light going green and the time of departure. As this study solely concerns the time the first pedestrian steps off the kerb before any others follow him, the origin of the perceived information is not social but is rather personal/private [25,26].

Globally, the decision making for a first pedestrian to cross, whether they do so at the green or at the red light, is influenced by the country of pedestrians but not by their gender. This does not mean that there is no effect of gender on the probability to cross the road at the red or the green light, but simply indicates that the index measured in this study does not reflect this effect. Indeed, we have already showed an effect of gender and country on risk-taking in a previous publication [13,26]. In Pelé et al. (2017), men crossed at the red light in 40.6% of cases, whilst women only did so in 25.7% of cases. The same difference was found in the present study but only in France and for the first process (before -400 $sec^{-100}$), since men and women in Japan showed an identical proportion of rule-breaking (about 2.2%). However, the number of data in each condition – i.e. crossing at the red or green light – is not taken into account in our survival analysis and our curve estimation. These analyses did however reveal that the threshold for crossing the road, whatever the light colour and the sensitivity, is not influenced significantly by the gender but is affected by the country variable. This might mean that fewer women cross at the red light than men but when they do so, they do it in the same way as men. The risk-taking is at two different levels here. Concerning the effect of the country, this factor affects not only the proportion of pedestrians crossing at the red light [13,26], but also the way they cross the road and their decision-making process. Although there is a lower number of Japanese pedestrians crossing at the red light, they have a lower threshold when they do so - meaning that the time they start to cross is closer to the moment the light turns green. However, the same shorter threshold was observed for Japanese pedestrians when crossing at the green light. In this condition, this means that either they are more concerned about watching the pedestrian light (private or personal information, see [26] for a discussion about this topic), or their motivation is higher than that of French pedestrians as they wait to cross at the green light. Indeed, waiting time influences the probability to cross at the red light but also the probability to cross faster and first at the green light [26]. We did not observe any difference for the sensitivity coefficient for gender or country, possibly because this process is more dependent on cognitive abilities that are not affected by the gender and/or the culture of the individual.

We also checked the influence of two other parameters on the probability to cross: the number of lanes and the number of pedestrians waiting at the time of crossing. Pedestrians tended to cross less when the road had four lanes compared to two lanes. We did not observe any rule-breaking for the site with six lanes. The effect of the number of lanes is amplified for Japanese pedestrians, whose probability to cross closer to the time of the light change is higher than that of French pedestrians. We also found a global effect of the number of pedestrians on the probability of crossing at the red light, with the number of pedestrians crossing at the red light decreasing as the number of waiting pedestrians increased. Whilst this effect is quite clear until about 11 pedestrians are present, a huge variation is then observed for numbers of pedestrians ranging from 11 to 40. This may be explained by two hypotheses. The first hypothesis is that, due to a decreasing number of observations per number of waiting pedestrians when the latter increase, we might have observed contrasted results (for one observation, the ratio is either -1 or 1), leading to this wide variation. The second hypothesis is that even if the probability of crossing at the red light per individual decreases with the number of pedestrians, the probability of observing one pedestrian crossing at the red light increases with the number of pedestrians. The higher the number of waiting persons, the more likely it is that one of these persons will prefer to cross at the red light. This hypothesis could also lead to wide variation when the number of waiting pedestrians increases.

This study highlights the social and environmental variables affecting the decision-making process in road crossing behaviours and for the first pedestrian to cross. Some components of the decision-making process, mainly risk sensitivity and temporal discounting, have already been identified for other behaviours [15,38,50,51]. Whatever the behaviours, these components are affected by both the country and the gender of the individual [16–18,21,52,52]. It is evident that these components cannot be controlled, except by better prevention and education about risk taking. However, these results show that certain other factors influence the probability of crossing at the red light, and it is possible to manipulate these factors to decrease risk taking and thus prevent accidents. Pedestrians do not like to wait for too long at the red light, and this increases their probability of crossing illegally. The duration of the pedestrian red light is important for road safety, and has to be limited [46,47]. Crossing behaviour is also influenced by the number of pedestrians waiting to cross. This "audience effect" [14,53] is known to have a strong social influence on human beings. It makes them comply and conform to the people surrounding them [27,54], and is strong in Japan [39]. [55] examined the effect of an image of a pair of eyes on contributions to an honesty box used to collect money for drinks in a coffee room. People paid nearly three times as much for their drinks when eyes were displayed compared to when other control images were displayed. We suggest the use of an image just above the pedestrian light showing the eyes of someone and indicating the risks of crossing at the red light. According to the study by Bateson et al. (2006), this should decrease the probability of red light crossing. The way pedestrians and traffic lights change also affects the probability that pedestrians will cross. The sequence of indications given by a traffic signal varies considerably between countries. The Austrian sequence of green–flashing green–amber–red–amber/red (Green) can, to our knowledge, only be found in Austria, Slovenia, Israel, Jordan and Cuba. Spain employs a green/amber indication instead of the flashing green, and a number of countries have abandoned the use of the amber/red combination (France, Italy, Belgium and Japan). [56] showed that the flashing green increases the number of early stops for cars and should reduce the number of accidents. However, according to our results, it could lead to increased numbers of pedestrians crossing at the red light. Countdowns for cars and pedestrians that are both visible to pedestrians could be a solution to decrease the number of illegal crossings and consequently reduce the number of accidents [57,58].

**Acknowledgements**

We thank Caroline Bellut, Elise Debergue, Charlotte Gauvin, Anne Jeanneret, Thibault Leclere, Lucie Nicolas, Florence Pontier and Diorne Zausa for their help in collecting data. We are grateful to Kunio Watanabe and Hanya Goro (Primate Research Institute, Kyoto University) for their help in obtaining autorisation for data collection. We thank Joanna Munro from Munro Language Services for the English language editing. Cédric Sueur was funded by the Chang Jiang Scholars Program at Sun-Yat Sen University to finalise this study.

**Author contributions statements**

MP and CS scored the data. MP, JLD and CS analysed the data. MP and CS wrote the manuscript. MP, JLD and CS reviewed the manuscript.

**Competing financial interests**

The author(s) declare no competing financial interests.

**References**

1. McNamara, J. M. & Houston, A. I. Integrating function and mechanism. *Trends Ecol. Evol.* **24,** 670–675 (2009).

2. Pelé, M. & Sueur, C. Decision-making theories: linking the disparate research areas of individual and collective cognition. *Anim. Cogn.* **16,** 543–556 (2013).

3. Bogacz, R. Optimal decision-making theories: linking neurobiology with behaviour. *Trends Cogn. Sci.* **11,** 118–125 (2007).

4. Marshall, J. A. R. *et al.* On optimal decision-making in brains and social insect colonies. *J. R. Soc. Interface* **6,** 1065–1074 (2009).

5. Laude, J. R., Stagner, J. P. & Zentall, T. R. Suboptimal choice by pigeons may result from the diminishing effect of nonreinforcement. *J. Exp. Psychol. Anim. Learn. Cogn.* **40,** 12–21 (2014).

6. Molet, M. *et al.* Decision making by humans in a behavioral task: Do humans, like pigeons, show suboptimal choice? *Learn. Behav.* **40,** 439–447 (2012).

7. Beck, J. & Forstmeier, W. Superstition and belief as inevitable by-products of an adaptive learning strategy. *Hum. Nat.* **18,** 35–46 (2007).

8. Cohen, L. J. Can human irrationality be experimentally demonstrated? *Behav. Brain Sci.* **4,** 317–331 (1981).

9. Martino, B. D., Kumaran, D., Seymour, B. & Dolan, R. J. Frames, Biases, and Rational Decision-Making in the Human Brain. *Science* **313,** 684–687 (2006).

10. Helbing, D. & Molnár, P. Social force model for pedestrian dynamics. *Phys. Rev. E* **51,** 4282–4286 (1995).

11. Moussaïd, M., Helbing, D. & Theraulaz, G. How simple rules determine pedestrian behavior and crowd disasters. *Proc. Natl. Acad. Sci.* **108,** 6884–6888 (2011).

12. Aoyagi, S., Hayashi, R. & Nagai, M. *Modeling of Pedestrian Behavior in Crossing Urban Road for Risk Prediction Driving Assistance System*. (SAE International, 2011).

13. Sueur, C., Class, B., Hamm, C., Meyer, X. & Pelé, M. Different risk thresholds in pedestrian road crossing behaviour: a comparison of French and Japanese approaches. *Accid. Anal. Prev.* **58,** 59–63 (2013).

14. Faralla, V., Innocenti, A. & Venturini, E. *Risk Taking and Social Exposure*. (Social Science Research Network, 2013).

15. Nagengast, A. J., Braun, D. A. & Wolpert, D. M. Risk-Sensitive Optimal Feedback Control Accounts for Sensorimotor Behavior under Uncertainty. *PLoS Comput Biol* **6,** e1000857 (2010).

16. Powell, M. & Ansic, D. Gender differences in risk behaviour in financial decision-making: An experimental analysis. *J. Econ. Psychol.* **18,** 605–628 (1997).

17. Tse, D. K., Lee, K., Vertinsky, I. & Wehrung, D. A. Does Culture Matter? A Cross-Cultural Study of Executives' Choice, Decisiveness, and Risk Adjustment in International Marketing. *J. Mark.* **52,** 81 (1988).

18. Wilson, M. & Daly, M. Competitiveness, risk taking, and violence: The young male syndrome. *Ethol. Sociobiol.* **6,** 59–73 (1985).

19. Pawar, D. S. & Patil, G. R. Pedestrian temporal and spatial gap acceptance at mid-block street crossing in developing world. *J. Safety Res.* **52,** 39–46 (2015).

20. Yannis, G., Papadimitriou, E. & Theofilatos, A. Pedestrian gap acceptance for mid-block street crossing. *Transp. Plan. Technol.* **36,** 450–462 (2013).

21. Holland, C. & Hill, R. The effect of age, gender and driver status on pedestrians' intentions to cross the road in risky situations. *Accid. Anal. Prev.* **39,** 224–237 (2007).

22. Tom, A. & Granié, M.-A. Gender differences in pedestrian rule compliance and visual search at signalized and unsignalized crossroads. *Accid. Anal. Prev.* **43,** 1794–1801 (2011).

23. Gallup, A. C. *et al.* Visual attention and the acquisition of information in human crowds. *Proc. Natl. Acad. Sci.* **109,** 7245–7250 (2012).

24. Webster, M. M. & Ward, A. J. W. Personality and social context. *Biol. Rev.* **86,** 759–773 (2011).

25. Faria, J. J., Krause, S. & Krause, J. Collective behavior in road crossing pedestrians: the role of social information. *Behav. Ecol.* **21,** 1236–1242 (2010).

26. Pelé, M. *et al.* Cultural influence of social information use in pedestrian road-crossing behaviours. *Open Sci.* **4,** 160739 (2017).

27. Cialdini, R. B. & Goldstein, N. J. Social influence: Compliance and conformity. *Annu Rev Psychol* **55,** 591–621 (2004).

28. Henrich, J. & Boyd, R. The Evolution of Conformist Transmission and the Emergence of Between-Group Differences. *Evol. Hum. Behav.* **19,** 215–241 (1998).

29. Rosenbloom, T. Crossing at a red light: Behaviour of individuals and groups. *Transp. Res. Part F Traffic Psychol. Behav.* **12,** 389–394 (2009).

30. Schmidt, S. & Färber, B. Pedestrians at the kerb–Recognising the action intentions of humans. *Transp. Res. Part F Traffic Psychol. Behav.* **12,** 300–310 (2009).

31. Czaczkes, T. J., Czaczkes, B., Iglhaut, C. & Heinze, J. Composite collective decision-making. *Proc R Soc B* **282,** 20142723 (2015).

32. Dyer, J. R. ., Johansson, A., Helbing, D., Couzin, I. D. & Krause, J. Leadership, consensus decision making and collective behaviour in humans. *Philos. Trans. R. Soc. B Biol. Sci.* **364,** 781–789 (2009).

33. Kurvers, R. H. J. M., Wolf, M., Naguib, M. & Krause, J. Self-organized flexible leadership promotes collective intelligence in human groups. *Open Sci.* **2,** 150222 (2015).

34. Fleming, T. R. & Harrington, D. P. *Counting Processes and Survival Analysis*. (John Wiley & Sons, 2011).

35. Klein, J. P. & Goel, P. K. *Survival Analysis: State of the Art*. (Springer, 1992).

36. Miller, R. G., Gong, G. & Muñoz, A. *Survival Analysis*. (1981).

37. Sueur, C. & Deneubourg, J.-L. Self-Organization in Primates: Understanding the Rules Underlying Collective Movements. *Int. J. Primatol.* **32,** 1413–1432 (2011).

38. Sueur, C. & Pelé, M. Risk should be objectively defined: reply to Zentall and Smith. *Anim. Cogn.* 1–3 (2015).

39. Benedict, R. *The chrysanthemum and the sword: patterns of Japanese culture*. (Houghton Mifflin Harcourt, 2005).

40. Fernandez, A. A. & Deneubourg, J. L. On following behaviour as a mechanism for collective movement. *J. Theor. Biol.* **284,** 7–15 (2011).

41. King, A. J. & Cowlishaw, G. Leaders, followers and group decision-making. *Integr. Commun. Biol.* **2,** 147–150 (2009).

42. Van Vugt, M., Hogan, R. & Kaiser, R. B. Leadership, followership, and evolution: Some lessons from the past. *Am. Psychol.* **63,** 182–196 (2008).

43. Sueur, C., Petit, O. & Deneubourg, J. Selective mimetism at departure in collective movements of Macaca tonkeana: an experimental and theoretical approach. *Anim. Behav.* **78,** 1087–1095 (2009).

44. García, L. V. Escaping the Bonferroni iron claw in ecological studies. *Oikos* **105,** 657–663 (2004).

45. Holm, S. A Simple Sequentially Rejective Multiple Test Procedure. *Scand. J. Stat.* **6,** 65–70 (1979).

46. Brosseau, M., Zangenehpour, S., Saunier, N. & Miranda-Moreno, L. The impact of waiting time and other factors on dangerous pedestrian crossings and violations at signalized intersections: A case study in Montreal. *Transp. Res. Part F Traffic Psychol. Behav.* **21,** 159–172 (2013).

47. Gårder, P. Pedestrian safety at traffic signals: a study carried out with the help of a traffic conflicts technique. *Accid. Anal. Prev.* **21,** 435–444 (1989).

48. Lachapelle, U. & Cloutier, M.-S. On the complexity of finishing a crossing on time: Elderly pedestrians, timing and cycling infrastructure. *Transp. Res. Part Policy Pract.* **96,** 54–63 (2017).

49. Amé, J.-M., Halloy, J., Rivault, C., Detrain, C. & Deneubourg, J. L. Collegial decision making based on social amplification leads to optimal group formation. *Proc. Natl. Acad. Sci. U. S. A.* **103,** 5835–5840 (2006).

50. Green, L. & Myerson, J. Exponential Versus Hyperbolic Discounting of Delayed Outcomes: Risk and Waiting Time. *Am. Zool.* **36,** 496–505 (1996).

51. Gould, J. P. Risk, stochastic preference, and the value of information. *J. Econ. Theory* **8,** 64–84 (1974).

52. Mihet, R. *Effects of Culture on Firm Risk-Taking: A Cross-Country and Cross-Industry Analysis*. (International Monetary Fund, 2012).

53. Zuberbühler, K. Audience effects. *Curr. Biol.* **18,** R189–R190 (2008).

54. Osman, L. M. Conformity or compliance? A study of sex differences in pedestrian behaviour. *Br. J. Soc. Psychol.* **21,** 19–21 (1982).

55. Bateson, M., Nettle, D. & Roberts, G. Cues of being watched enhance cooperation in a real-world setting. *Biol. Lett.* **2,** 412–414 (2006).

56. Köll, H., Bader, M. & Axhausen, K. W. Driver behaviour during flashing green before amber: a comparative study. *Accid. Anal. Prev.* **36,** 273–280 (2004).

57. Keegan, O. & O'Mahony, M. Modifying pedestrian behaviour. *Transp. Res. Part Policy Pract.* **37,** 889–901 (2003).

58. Lipovac, K., Vujanic, M., Maric, B. & Nesic, M. The influence of a pedestrian countdown display on pedestrian behavior at signalized pedestrian crossings. *Transp. Res. Part F Traffic Psychol. Behav.* **20,** 121–134 (2013).

**Tables**

Table 1: Information about the studied sites in France and in Japan. Road-crossing speed was estimated by scoring the crossing speed of 20 random pedestrians for each site.

|  | France - Strasbourg | | |  |
|---|---|---|---|---|
| Sites | Train Station | Pont des Corbeaux | Place Broglie |  |
| Coordinates | 48.584474, 7.736135 | 48.579509, 7.750745 | 48.584559, 7.748628 |  |
| Lanes | 2*1 | 2*2 | 2*1 |  |
| Mean pedestrian flow per hour | 667 | 612 | 850 |  |
| Mean road crossing speed | 0.96±0.05 | 1.11±0.29 | 1.01±0.16 |  |
| Dates of scoring | 02/07-07/07/2014 | 01/10-25/10/2014 | 15/02-09/03/2015 |  |
|  | Japan - Nagoya | | | |
| Sites | Train Station | Maruei | Excelco | Osu-Kannon |
| Coordinates | 35.170824, 136.884328 | 35.168638, 136.905740 | 35.166891, 136.907284 | 35.159316, 136.901697 |
| Lanes | 2*3 | 1*1 | 2*1 | 2*1 |
| Mean pedestrian flow per hour | 480 | 645 | 869 | 814 |
| Mean road crossing speed | 1.10±0.22 | 1.15±0.21 | 0.98±0.21 | 1.07±0.18 |
| Dates of scoring | 13/06-05/07/2011 |  | 27/01-05/02/2015 | |

Table 2: Values of threshold S and sensitivity q from the sigmoid curves fitting with the observed data for each category (country-gender) as well as statistical values. All p-values are < 0.00001

|  |  | France - man | France - woman | Japan - man | Japan - woman |
|---|---|---|---|---|---|
| **Red light** | **Threshold S** | -400 | -170 | Non-applicable due to small dataset (two points per condition) | |
| <-400 | **Sensitivity q** | 2.86 | 5.56 | | |
|  | **R²** | 0.98 | 0.98 | | |
|  | **F** | 1882 | 2242 | | |
| **Before light turns green** | **Threshold S** | -230 | -100 | -60 | -60 |
| [-400; 0[ | **Sensitivity q** | 2.5 | 2.2 | 2 | 2 |
|  | **R²** | 0.99 | 0.97 | 0.98 | 0.98 |
|  | **F** | 3196 | 571 | 1656 | 734 |
| **green light** | **Threshold S** | 80 | 90 | 70 | 70 |
| ≤0 | **Sensitivity q** | 2.5 | 3 | 3 | 3.5 |
|   | R² | 0.99 | 0.99 | 0.98 | 0.99  |
|   | F  | 1727 | 1319 | 6902 | 14890 |

Table 3: Equation and statistical values of survival curves for departure time according to the country and the number of lanes. The six lane condition in Japan does not appear here as no illegal crossings were observed. All p-values are < 0.00001

| Country | Number of lanes | Equation ($y=a*e^{-bx}$) | Log gradient -(1/b) | R² | F |
|---|---|---|---|---|---|
| France | 2 | $y=0.954*e^{-0.0006x}$ | -1666.67 | 0.99 | 8032 |
| Japan | 2 | $y=0.9*e^{-0.009x}$ | -1000 | 0.82 | 67 |
| France | 4 | $y=1.0012*e^{-0.003x}$ | -333.33 | 0.99 | 3128 |
| Japan | 4 | $y=1.340*e^{-0.007x}$ | -142.86 | 0.90 | 269 |

Table 4: Equation and statistical values of survival curves for waiting time according to the country and the gender of crossing pedestrians. All p-values are < 0.00001

| Country | Gender | Equation ($y=a*e^{-bx}$) | Log gradient (1/b) | R² | F |
|---|---|---|---|---|---|
| France | Man | $y=0.852*e^{-0.0008x}$ | 1250 | 0.98 | 3663 |
| Japan | Man | $y=1.143*e^{-0.0003x}$ | 3333.3 | 0.96 | 620 |
| France | Woman | $y=0.942*e^{-0.0006x}$ | 1666.7 | 0.99 | 5556 |
| Japan | Woman | $y=1.048*e^{-0.0003x}$ | 3333.3 | 0.94 | 220 |

**Figure legends:**

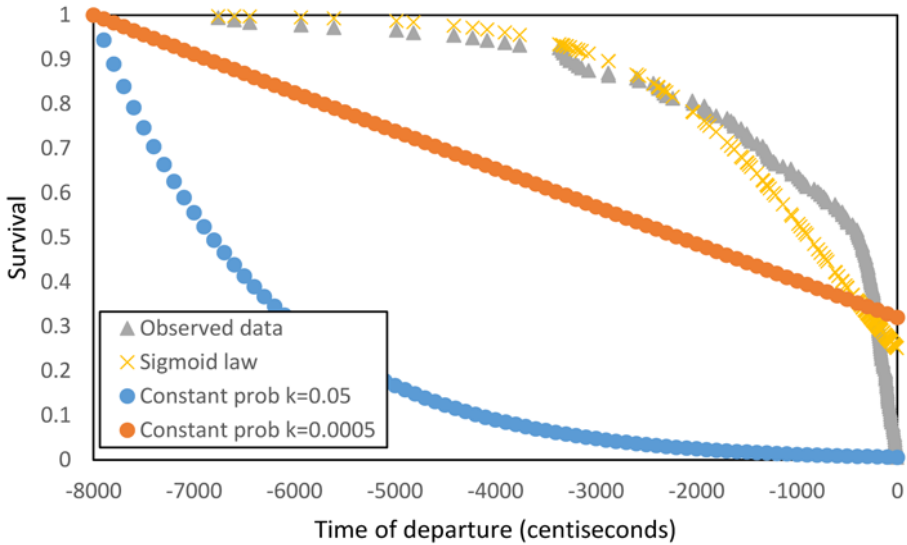

Figure 1: a. Survival analysis of departure time for the total dataset, i.e., at the red light (negative values) and at the green light (positive value) for the observed data (grey) and for the theoretical data (yellow). b. Survival values of theoretical data according to survival values of observed data. The thin black line represents the correlation we should observe between the two survivals if the rule underlying the departure of a pedestrian follows only one rule (here a sigmoid curve, equation 1). For Fig.1 a., 0 indicates the time at which the light turns green.

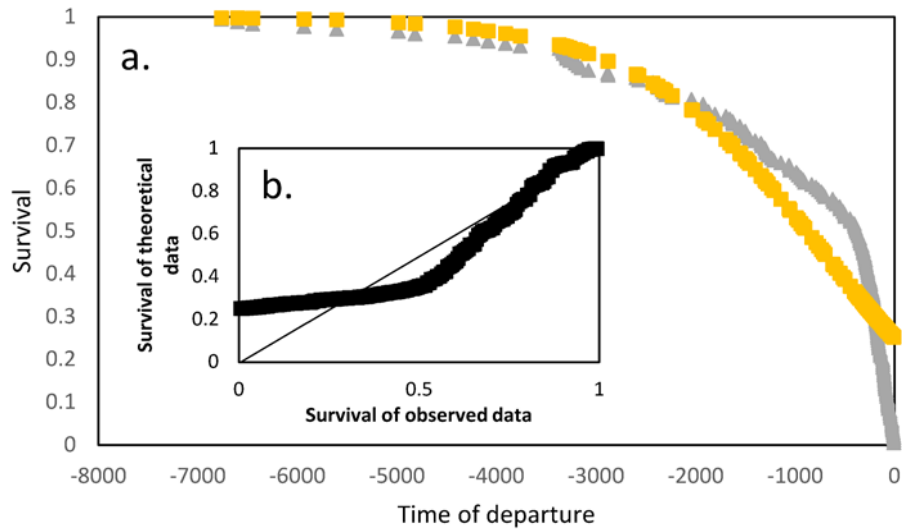

Figure 2: a. Survival analysis of departure time at the red light for the observed data (grey). Yellow points indicate theoretical survival following a sigmoid law (non constant probability of departures). Orange and blue points indicate theoretical survival following contact probability of departure (blue: k=0.05, high probability to cross at the red light; orange: k=0.0005, low probability to cross at the red light).

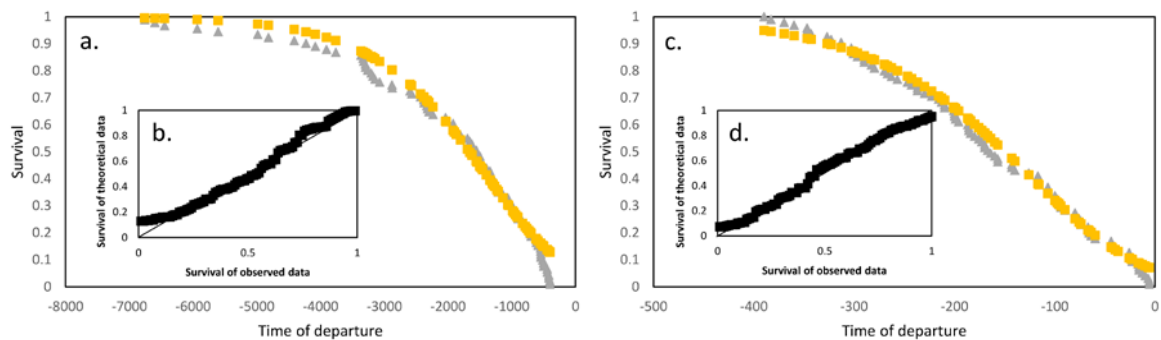

Figure 3: Survival analysis of departure time at the red light for the observed data (grey) and for the theoretical data (yellow) (a.) before -400 sec-100 and (b.) between -400 and 0 sec-100. Survival curves of theoretical data compared to survival of observed data (a.) before -400centisconds and (b.) between -400 and 0 sec-100. The thin black line represents the correlation we should observe between the two survivals if the rule underlying the departure of a pedestrian follows only one rule (here, a sigmoid curve, Equation 1). For 3 a. and c., 0 indicates the time at which the light turns green.

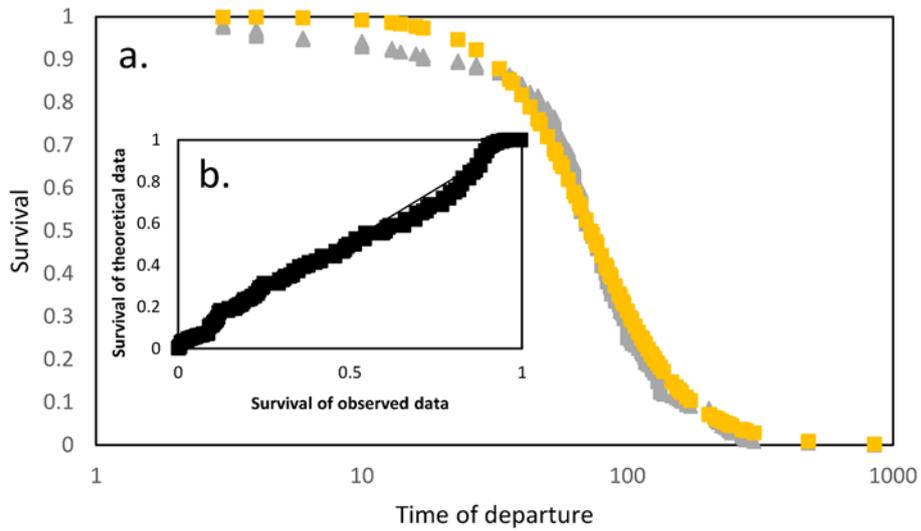

Figure 4: a. Survival analysis of departure time (as log. scale) at the green light for the observed data (grey) and for the theoretical data (yellow). b. Survival values of theoretical data compared to survival values of observed data. The thin black line represents the correlation we should observe between the two survival values if the rule underlying the departure of a pedestrian follows only one rule (here, a sigmoid curve, Equation 1). For 4a., 0 (1; as log scale) indicates the time at which the light turns green.

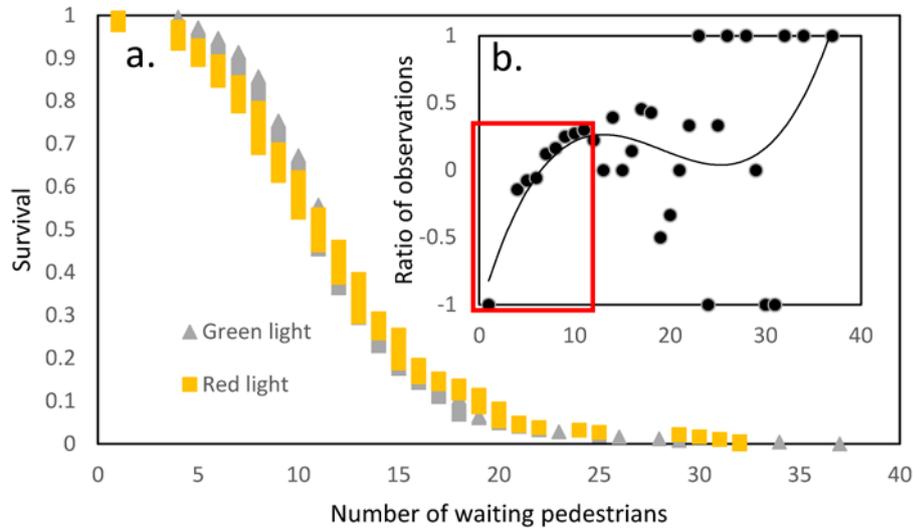

Figure 5: a. Survival analysis of the number of waiting pedestrians at the time of crossing at the green light (grey) and the red light (yellow). b. Ratio of observations for crossings according to the number of waiting pedestrians. -1 indicates that all crossings were made at the red light. 1 indicates that all crossings were made at the green light. 0 means that half of the observations were made at the green light (and, of course, half at the red light). A red square indicates data that can only be significantly explained ($P<0.05$).